\documentstyle[12pt]{article}
\begin{document}
\pagestyle{empty}
\setlength{\oddsidemargin}{0.5cm}
\setlength{\evensidemargin}{0.5cm}
\setlength{\footskip}{2.0cm}
\newcommand{\be}{\begin{eqnarray}}
\newcommand{\en}{\end{eqnarray}}
\renewcommand{\thepage}{-- \arabic{page} --}
\newcommand{\csl}{{\sl c}} \newcommand{\ssl}{{\sl s}}
\newcommand{\dr}{{\mit\Delta}r}
\newcommand{\drmt}{{\mit\Delta}r\lbrack m_t \rbrack}
\newcommand{\Born}{\lbrack{\rm Born}\rbrack}
\def\R#1{$\lbrack #1 \rbrack$}
\def\mib#1{\mbox{\boldmath $#1$}}
\def\nmib#1{\mbox{\normalsize\boldmath $#1$}}
\def\smib#1{\mbox{\small\boldmath $#1$}}
\def\rsr{\buildrel <\over\sim}
\def\lsl{\buildrel >\over\sim}
\def\lbo#1#2{ {\lower#2ex\hbox{$#1$} } }
\def\rs{\lbo{\rsr}{0.5}  }
\def\ls{\lbo{\lsl}{0.5}  }
%
\vspace*{-2cm}\noindent
\hspace*{10.6cm} TOKUSHIMA 94-02 \\
\hspace*{10.6cm} YCCP-9404 \\
\hspace*{10.6cm} June 1994 \\

\vspace*{1cm}

\centerline{\large{\bf Is the Standard Electroweak Theory Happy}}

\vskip 0.15cm
\centerline{\large{\bf with $\!\mib{m}_{\nmib{t}}\sim$ 174 GeV?}}

\vspace*{1.5cm}

\renewcommand{\thefootnote}{\arabic{footnote})}
\centerline{\sc \phantom{*)}Zenr\=o HIOKI\,$^{\rm a),}\!$
\footnote{E-mail: hioki@ias.tokushima-u.ac.jp, hioki@jpnyitp,
a52071@jpnkudpc}
\ {\rm and}\ Ryuichi NAJIMA$^{\rm b),}\!$
\footnote{E-mail: najima@jpnkekvax}}

\vspace*{1.5cm}

\centerline{a)~{\sl Institute of Theoretical Physics,~
University of Tokushima}}

\centerline{\sl Tokushima 770,~JAPAN}

\vskip 0.3cm
\centerline{b)~{\sl Department of General Education,~
Yokohama College of Commerce}}

\centerline{\sl Yokohama 230,~JAPAN}

\vspace*{2.4cm}

\centerline{ABSTRACT}

\vspace*{0.4cm}
\baselineskip=20pt plus 0.1pt minus 0.1pt
Based on the recent CDF report on the top-quark, we have carried
out an analysis on the Higgs mass within the minimal standard
electroweak theory using the latest data on the $W$-mass.
Although this theory is in quite a happy situation now, we wish
to point out that more precise measurements of $M_W$ and $m_t$ in
the future are crucial and they could come to require some new
physics beyond it.

\vfill
\newpage
\pagestyle{plain}
\renewcommand{\thefootnote}{\sharp\arabic{footnote}}
\setcounter{footnote}{0}
\baselineskip=21.0pt plus 0.2pt minus 0.1pt

Recently, CDF collaboration at Fermilab Tevatron collider has
reported evidence of top-quark pair productions \cite{Top}. There
its mass has been estimated to be $m_t^{exp}=174\pm 16$ GeV. Its
final establishment must come after D0 collaboration confirms their
results, but this observation will surely work as a new strong
experimental support to the minimal standard electroweak theory
with three fermion generations (the electroweak theory, hereafter).
It is also noteworthy that very heavy top ($\sim$ 160-180 GeV) has
already been anticipated through analyses of low- and high-energy
precision electroweak data \cite{Pretop} before the above CDF
report.

It seems that the electroweak theory is in a very happy situation.
This is true at present, but one might feel that the above
$m_t^{exp}$ is a little too heavy. In this short note, we have
studied this problem briefly. As a result, we wish to point out
that more precise determinations of $m_t$ and $M_W$ might
bring us into another very stimulating situation. The important
point is the $m_{\phi}$(the Higgs-boson mass)-dependence of the
$M_W$-$M_Z$ relation derived from the $\mu$-decay in the electroweak
theory. We use here the $M_W$-$M_Z$ formula given in \cite{Hio},
which has already been confirmed to be consistent with other
calculations \cite{Sir}.

We start our discussion with summarizing phenomenological analyses
on the Higgs mass. Ellis et al. obtained $m_{\phi}<$ 250 GeV at 95
\% C.L. independently of $m_t$ \cite{EFL}. The results by Novikov
et al. in \cite{Pretop} and by Jacobsen \cite{Jac} are both not so
drastic, but still low $m_{\phi}$ is favored and $1\sigma$ region
gives an upper bound $m_{\phi}\ \rs$ 200-300 GeV. (In the latter
analyses, the recent SLD measurement of $\sin^2\theta_W^{eff}$
\cite{SLD} is also used.)

However, this does not mean that all the electroweak quantities
used there demand low-mass Higgs boson. Indeed, the central value
of $M_W^{exp}$ ($M_W^{exp}=80.21\pm 0.18$ GeV by UA2+CDF+D0
\cite{Sal}) and that of $m_t^{exp}$(=174 GeV) require very
heavy Higgs ($\sim$ 1.7 TeV) via the well-known relation
\be
M_W^2={1\over 2}M_Z^2
\biggl\{ 1+
\sqrt{\smash{1-{{2\sqrt{2}\pi\alpha}\over{M_Z^2 G_F (1-\dr)}}}
\vphantom{A^2\over A}
}~\biggr\}, \label{eqaa}
\en
where $\alpha=1/137.036$, $G_F=1.16639\times 10^{-5}$ GeV$^{-2}$,
$M_Z=91.1899\pm 0.0044$ GeV \cite{Cla}, and $\dr$ is the one-loop
corrections to the $\mu$-decay amplitude.\footnote{In actual
    calculations, $m_t^2$ term resummation \cite{CHJ} plus QCD
    corrections to the top-quark loop \cite{HKl} have been taken
    into account in addition to Eq.(\ref{eqaa}).}\ 
At present, it does not cause any serious trouble since $m_{\phi}$
as low as 80 GeV is also allowed if we take into account
${\mit\Delta}m_t^{exp}=\pm 16$ GeV and ${\mit\Delta}M_W^{exp}
=\pm 0.18$ GeV.\footnote{$M_W-M_W^{exp}=0.22\pm 0.21$ GeV and $0.21
    \pm 0.21$ GeV for $m_{\phi}=70$ GeV and 80 GeV respectively.}\
That is, the $m_{\phi}$-dependence of the $M_W$-$M_Z$ relation is
not strong. That is why $\chi^2$ takes its minimum at low $m_{\phi}$
even when $M_W^{exp}$ is taken into account in an analysis.

When LEP II starts, the $W$-mass is expected to be determined
very precisely: ${\mit\Delta}M_W^{exp}\sim\pm 0.05$ GeV \cite{Kaw}.
We may also expect that $m_t$ will eventually be measured with
better precision. We assume here tentatively that
${\mit\Delta}m_t^{exp}\sim\pm 5$ GeV will be possible in the
near future. In this case, a constraint from the $W$-mass
becomes much stronger. Concretely, ${\mit\Delta}m_t^{exp}=\pm
5$ GeV produces an error of $\pm 0.03$ GeV in the $W$-mass
calculation. Combining this with ${\mit\Delta}M_W^{exp}=\pm
0.05$ GeV and a theoretical ambiguity ${\mit\Delta}M_W=\pm 0.03$
GeV (which has been a bit overestimated for safety), we can
compute $M_W-M_W^{exp}$ with an error of about $\pm 0.07$ GeV.
As an example, let us assume that the central values of $M_W^{exp}$
and $m_t^{exp}$ do not change. Then, $M_W-M_W^{exp}$ becomes $0.13
\pm 0.07$ GeV for $m_{\phi}=300$ GeV. It means that $m_{\phi}=300$
GeV is ruled out at $1.9\sigma$ level within the minimal standard
electroweak theory. Similarly, even $m_{\phi}=600$ GeV is not allowed
though at $1.1\sigma$ level ($M_W-M_W^{exp}=0.08\pm 0.07$ GeV). To be
consistent with the data at $1\sigma$ level, $m_{\phi}$ has to be
at least 650 GeV.

On the other hand, it is obvious that the upper bound on $m_{\phi}$
derived in analyses without $M_W$ becomes lower than the one in
those with $M_W$ since $M_W^{exp}$ itself favors high mass Higgs.
This means that we are led to another very exciting situation:
$M_W^{exp}$ demands heavy Higgs: $m_{\phi}\ \ls$ 650 GeV, while
the others need $m_{\phi}\ \rs$ 200-300 GeV. As already mentioned,
the central values of $M_W^{exp}$ and $m_t^{exp}$ demand $m_{\phi}
\sim$ 1.7 TeV. Even if we limit discussions to perturbation
calculations, such extremely heavy Higgs will cause serious problems
\cite{DM} (see also \cite{Ghi} and references cited therein).

It will be difficult to present this conclusion more strongly, e.g.,
at $3\sigma$
due to the well-known fact that low energy quantities do not have
$m_{\phi}^2$ terms at one-loop order \cite{Vel}. Nevertheless, if a
situation like that comes to be real, it must be quite interesting,
and we may need to consider some new physics beyond the standard
electroweak theory which makes opposite contribution to $M_W$ and
the other quantities. Precise measurements of $M_W$ and $m_t$ are
therefore considerably significant.

\centerline{ACKNOWLEDGEMENTS}

\vspace*{0.3cm}
We would like to thank T. Kawamoto, T. Kobayashi, T. Kon, K. Kondo,
K. Miyabayashi and A. Miyamoto
for kind correspondences, and the theory division of Institute for
Nuclear Studies (INS), University of Tokyo for giving us warm
hospitality, where a part of this work was done. INS and KEK
computers were used for a part of our calculations.

\vskip 0.8cm

\end{document}